\documentclass{ws-procs9x6}
\usepackage{slashed} 
\begin{document}

\title{LORENTZ VIOLATION BY QUARK CONDENSATION}

\author{CHI XIONG}

\address{Department of Physics, University of Virginia\\
Charlottesville, VA 22904-4714, USA\\
E-mail: cx4d@virginia.edu}

\begin{abstract}
At the TeV scale, heavy quarks, for example the 4th generation quarks of the Standard Model with four generations, can form condensates which dynamically break the electroweak symmetry. A dense quark system may form other types of condensates which dynamically break the Lorentz symmetry. These condensates are described by a Nambu-Jona-Lasinio type effective action, 
similar to the quark condensation models in hadronic matter with finite density. 
The vacua corresponding to these two types of condensates compete for the global minimum of the effective potential,
depending on the energy scale and the related strong dynamics. The resultant Goldstone gauge boson may produce 
observable effects in relativistic heavy-ion colliders.

\end{abstract}

\bodymatter

\section{Quark condensates and electroweak symmetry breaking}

\begin{figure}[!tbp]
\centering
    \psfig{file=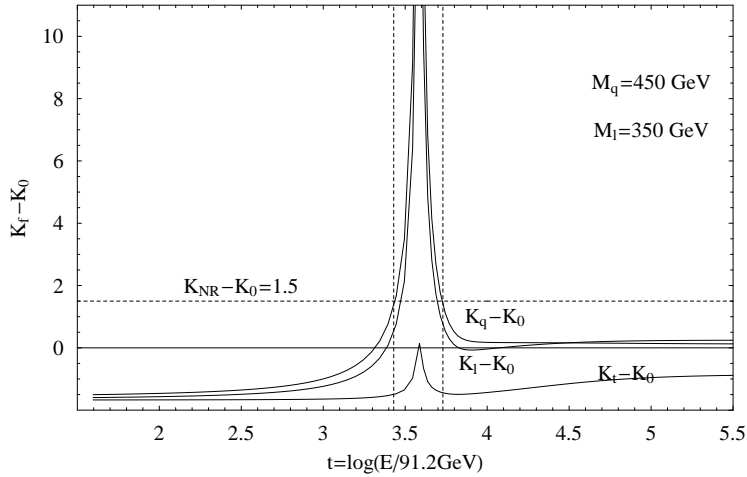,width=4.0in} \\
\caption{{\small ($K_0=1.68$) $K_q, K_l$ are for the 4th generation quarks and leptons respectively and $K_t$ for the top quark. 
The horizontal dotted line indicates an estimate of $K_f$ where the nonrelativistic method is still applicable and the vertical dotted lines enclose the region where a fully relativistic approach is needed. It shows that the 4th generation can
easily form strong bound states while the top-quark (the lowest curve) hardly can.}}
\label{condensates}
\end{figure}

The Higgs sector of the Standard Model might have a Landau-Ginzburg effective description, in which the Higgs doublet
is a condensate of fermion and antifermion. The constituent fermions could be the top quark in the top-quark condensation
models, or technifermions in the technicolor models (see Refs.\ \refcite{SD} for reviews). A new interaction, e.g., a four-fermion interaction or technicolor, is introduced at the composite scale. 
Recently it has been shown\cite{HX12} that in the Standard Model with four generations (SM4), the renormalization group 
evolutions of the Yukawa and quartic couplings can reach a quasi-fixed point at the TeV scale ($\Lambda_{FP}$), where
the Yukawa couplings become strong enough for the 4th generation to form condensates. 
This can be seen by considering a nonrelativistic Higgs-exchange potential between a fermion and an antifermion
\begin{equation}
V(r) = - \alpha_Y e^{-m_H(r) r} /r .
\end{equation}
By numerical analysis, the bound state condition is found to be $K_f>1.68$, where $K_f$
is expressed in terms of the Yukawa couplings and the quartic couplings as
$K_f = g_f^3 / (16 \pi \sqrt{\lambda})$.
The values $ K_f - K_0$ for the 4th generation and for the top quark are plotted against the energy scale
in Fig.\ \ref{condensates}. 

To study the strong dynamics of Yukawa couplings in a relativistic way,
the Schwinger-Dyson equation is used to find the critical coupling for the Yukawa interactions\cite{HX3}, following the
strong QED case considered by Leung, Love, and Bardeen\cite{LLB}. It also leads to the condensates of the
4th generation at the TeV scale, consistent with the renormalization group and the nonrelativistic analyses. If the 4th generation
really exists, the masses of the 4th generation quarks should be greater than 400 GeV, in order to have $\Lambda_{FP}$
located at the TeV scale. We will need such a cutoff scale in the next section.

\section{Dynamical Lorentz symmetry breaking }

The condensate $\langle \bar{\psi} \psi \rangle$ can be considered as a composite scalar and one
may ask: what about a composite vector? 
The possibility of the bilinear form $\bar{\psi} \gamma_{\mu} \psi $ developing a nonvanishing vev 
has been considered by Bjorken\cite{BJ} and others\cite{AK, Kraus, Jenkins,BK}. 
Some realistic models are also built in the quark systems with finite density, for example, Langfeld, Reinhardt,
and Rho (LRR) found that\cite{LRR} in the dense hadronic medium, 
there exists a critical chemical potential $\mu_c$, above which the system can have spontaneous
Lorentz symmetry breaking (on top of an explicit breaking). This new phase is used to explain
the enhanced dilepton production in relativistic heavy-ion collisions. In this section we apply
the LRR mechanism to the physics at the TeV scale, 
combine these models in the SM4 scenario, and focus on the 4th generation quark ($t'$), although $t'$ might be other
heavy fermions beyond the Standard Model, such as technifermions in technicolor models, 
fermionic Kaluza-Klein excitations in extra-dimensional models, gauginos in supersymmetric models, etc. 
The most important property that we look for is if two phases below `compete'
for either external (spacetime) or internal symmetry breaking: 
\begin{equation} \label{2phase}
\textrm{Phase I}: \langle \bar{t'} \gamma_0 t' \rangle = 0, ~\langle \bar{t'} t' \rangle \neq 0, ~~
\textrm{Phase II}: \langle \bar{t'} \gamma_0 t' \rangle \neq 0, ~\langle \bar{t'} t' \rangle = 0. 
\end{equation}
Phase I is an electroweak broken but Lorentz invariant state, while in phase II the Lorentz symmetry
is broken but the electroweak symmetry is restored. 
For simplicity we only consider a truncation of the SM4 to one flavor $ t' $. 
Our toy model is described by a Nambu-Jona-Lasinio type Lagrangian from\cite{LRR}
\begin{eqnarray} \label{LRR2}
  \mathcal{L}_{\textrm{NJL}} &=& 
\mathcal{L}_{0} + \mathcal{L}_{s}
+ \mathcal{L}_{\textrm{int}} - \mathcal{L}_{s}, \\ 
  \mathcal{L}_{0} &=& 
i \bar{t'}{\slashed \partial} t', 
\qquad \mathcal{L}_{s}=\bar{t'} ( - m - \mu \gamma^0 ) t', \\ 
  \mathcal{L}_{\textrm{int}} &=& 
\frac{1}{2 \lambda_s} (\bar{t'} t' )^2 
+ \frac{1}{2 \lambda_v} (\bar{t'} \gamma^{\mu} t')^2,
\end{eqnarray}
where $\lambda_s = N m_s^2, \lambda_v = N m_v^2 $ and $ N$ is the number of colors.
Equation \eqref{LRR2} can be obtained as an effective theory from
\begin{equation} 
\label{LRR1}
 \mathcal{L} = \bar{t'} ( i {\slashed \partial} - \sigma + i {\slashed A} ) t' 
 - \frac{\lambda_s}{2} (\sigma - m)^2 - \frac{\lambda_v}{2} [(A_0 - \mu)^2 + A^2_i] + \cdots
\end{equation}
by integrating out the fields $\sigma $ and $A_{\mu} $, where 
$A_{\mu} $ is Bjorken's vector field and $ \sigma $ comes from the truncation of
the `Higgs' field $\Phi$ parametrized by $\Phi = \sigma + i \tau^a \pi^a $. 
The $\pi^a$-related terms and other kinetic terms are included in the ellipsis of Eq.\ \eqref{LRR1}. 
In the large $N$ expansion we can use the mean field approximation at the leading order. 
Following the standard procedure, one obtains an effective potential\cite{LRR}
\begin{eqnarray} \label{Ueff}
\nonumber
 \frac{1}{N} U_{\textrm{eff}} (\sigma, A_{\mu})
 &=& \frac{m_s^2}{2} (\sigma - m)^2 + \frac{m_v^2}{2} [(A_0 - \mu)^2 + A^2_i] \\ 
 && - \frac{1}{N \Omega_4} \textrm{Tr} \ln \{ i {\slashed \partial} - \sigma + i {\slashed A} \} + \mathcal{O}(\frac{1}{N}),
\end{eqnarray}
where the trace term is
\begin{equation}
\frac{\Omega_4}{8 \pi^3} \int^{1}_{-1} dt \sqrt{1-t^2} \int^{\Lambda^2}_0 du ~u \ln [ (u + \sigma^2 - A^2) + 4 u t^2 A^2 ] 
\end{equation}
with $A^2 = A_{\mu} A_{\mu} $ and $\Omega_4$ is the 4-dimensional Euclidean space volume.
In Ref.\ \refcite{LRR} the cutoff $\Lambda$ satisfies $ \Lambda^2 = 8 \pi^3 m_s^2 $.
This relation is quite natural in the SM4 scenario, where
the cutoff $\Lambda$ is taken to be the SM4 quasi-fixed point scale, i.e. $ \Lambda =\Lambda_{FP} \sim $TeV, 
and $m_s$ is supposed to be the order of $10^2$ GeV. 
The extrema of the effective potential $U_{\textrm{eff}} $ yields the gap equation which can
also be obtained from the Schwinger-Dyson equation based on Eq.\ \eqref{LRR2}. 

Does the effective potential \eqref{Ueff} allow the phase transition between those two states in Eq.\ \eqref{2phase}?
In fact Langfeld, Reinhardt, and Rho have already found\cite{LRR} that such a phase transition is possible provided
that the chemical potential $\mu$ exceeds a critical value $\mu_c$. Here we have a similar mathematical structure, 
however the physical scale is $\Lambda_{FP} $ instead of $\Lambda_{QCD}$, and the internal symmetry is the
electroweak symmetry instead of the chiral symmetry. 
It is probably easier to understand this scaled-up mechanism if we consider the technicolor
theories as a scaled-up version of QCD. 
If $t'$ does not represent the 4th generation quark but 
a technifermion instead, one can readily incorporate the LRR mechanism into 
the technicolor scenario at the scale of $\Lambda_{TC}$ and study the `technihadrons' 
for the electroweak symmetry breaking. The effective
theory below $\Lambda_{TC}$ might as well be described by the same type Nambu-Jona-Lasinio effective action.

\section{Discussion}

If the Lorentz violation is determined only by dynamical symmetry breaking, one then
expects that a massless vector appears, e.g., Bjorken's emergent gauge boson (for a systematic study
of spontaneous Lorentz violation, Nambu-Goldstone modes and gravity, see Ref.\ \refcite{BK}). 
The diquark condensates can be dynamically generated without triggers, as long as the attraction
between quarks becomes strong enough at some energy scale and the resultant vacuum is the global
minimum. The LRR mechanism leads to another
possibility: Eq.\ \eqref{LRR1} can be interpreted as the spontaneous Lorentz symmetry breaking induced by an explicit 
breaking, since one can always split $\mu=\mu_c + \mu' $ if $\mu > \mu_c$ and consider the $\mu_c$ term as the
source for the explicit Lorentz violation. 
In this case the Goldstone boson is `light' (compared with
the condensation scale) but not massless, similar to the pion mass in the chiral symmetry breaking. 
Note that this kind of Lorentz violation cannot be `gauged' away as in some theories with ostensibly explicit Lorentz violations,
since it can modify the vacuum structure and induce a spontaneous Lorentz violation. 
It could appear as a new gauge boson and be phenomenologically observable at the TeV scale. In Refs.\ \refcite{Clark} 
the dynamics of a massive vector and its couplings to the SM are studied from the extra-dimension point of view: 
a new vector field, originally a part of the higher dimensional metric, becomes massive after absorbing the brane fluctuations.
Such massive vector also appears in the gauged ghost-condensation models\cite{Luty}. 
If the Goldstone gauge boson from spontaneous Lorentz symmetry breaking is a condensate of some heavy fermions, its
properties may be further investigated by studying the Bethe-Salpter equation.

\section*{Acknowledgments}

I would like to thank P.Q. Hung for fruitful conversations on the Standard Model with four generations. 
I also thank J. Bjorken, S.T. Love, and A. Kosteleck\'y for valuable discussions.
This work is supported by the US Department of Energy under grant No. DE-FG02-97ER41027.

\end{document}